# DELIVERY RING LATTICE MODIFICATIONS FOR TRANSITIONLESS DECELERATION*

J. A. Johnstone[†], M. J. Syphers[‡], Fermilab, Batavia, United States
[‡]also at Northern Illinois University, DeKalb, United States

*Abstract*

A portion of the remnant Tevatron program infrastructure at Fermilab is being reconfigured to be used for the generation and delivery of proton and muon beams for new high-precision particle physics experiments. With the 8 GeV Booster as its primary source, the Mu2e experiment will receive 8.9 GeV/c bunched beam on target, after being stored and slow spilled from the Delivery Ring (DR) -- a refurbished debuncher ring from Tevatron antiproton production. For the Muon g-2 experiment, the DR will be tuned for 3.1 GeV/c to capture muons off of a target before sending them to this experiment's Storage Ring. The apertures in the beam transport systems are optimized for the large muon beams of this lower-energy experiment. In order to provide further flexibility in the operation of the DR for future possible low-energy, high-intensity particle physics experiments (REDTOP[1], for example) and detector development, investigations are underway into the feasibility of decelerating beams from its maximum kinetic energy of 8 GeV level to lower energies, down to 1-2 GeV. In this paper we look at possible lattice modifications to the DR to avoid a transition crossing during the deceleration process. Hardware requirements and other operational implications of this scheme will also be discussed.

## OVERVIEW

The DR has a 505 m circumference comprised of 6 arcs and 6 straight sections (with dihexal symmetry). Natural transition energy of the machine is at $\gamma_t$ = 7.64 (6.23 GeV kinetic energy, or 7.58 GeV/c momentum). Large beam losses are expected during deceleration unless transition crossing issues are addressed. This note outlines a $\gamma_t$ scheme in which beam momentum is always below transition gamma ($\gamma_t$ always boosted above the beam $\gamma$).

The system consists of 18 quadrupoles arranged in 6 groups of triplets. Each triplet has 2 quads in the arcs ($n$Q13 and $n$Q19, $n$ = 1→6) and a double-strength quad at the end of each straight section ($n$Q07). Each quad is separated from its neighbor in the triplet by 180° of phase advance. With this arrangement optical perturbations are localized within each triplet. The 2 arc quads in each triplet reduce the average dispersion, while the double strength quads cancel the beta wave generated by the pairs of quads in the arcs. Straight section optics are thereby unaffected so the nominal injection tune of the M3 beamline is valid.

* Work supported by Fermi Research Alliance, LLC under contract no. DE-AC02-07CH11359
† jjohnstone@fnal.gov
‡ syphers@fnal.gov, msyphers@niu.edu

## VARIABLE MOMENTUM COMPACTION

At the injection energy of 8 GeV ($p$ = 8.889 GeV/c) the beam Lorentz factor $\gamma$ = 9.53. The proposal here is to use the arc $\gamma_t$ quads to increase $\gamma_t$ to 10.03 prior to injection and maintain $\gamma_t$ at 0.5 units above beam $\gamma$ through deceleration until $\gamma_t$ = 7.64 and beam $\gamma$ = 7.14 ($p$ = 6.633 GeV/c). From that point on the DR lattice corresponds to the nominal design configuration.

Transition gamma is related to the momentum compaction factor $\alpha$ via:

$$\gamma_t = \sqrt{1/\alpha} \quad (1)$$

where $\alpha$ is the average of the dispersion divided by the radius of curvature around the ring:

$$\alpha = \langle \tfrac{D}{\rho} \rangle \quad (2)$$

Increasing $\gamma_t$ therefore requires that the average dispersion be reduced. The arc quads in each triplet have an integrated field $q$:

$$q \equiv B'l/B\rho \quad (3)$$

where at 8 GeV $B\rho$ = 29.650 T·m and for an SQC style quadrupole $l$ = 0.6668 m.

If $m$ FODO cells span 180° of phase advance then, in thin lens approximation, at the F quads separating a pair of $q$'s the maximum, matched, dispersion $D$ is reduced by the factor:

$$D_{max} \rightarrow D_{max} \cdot (1 - q \cdot \beta_{max} \cdot \sin(n/m\,\pi)) \quad (4)$$

with $n$ = 1 → ($m$-1). $D$ is reduced at each of the F quads intervening between a pair of $q$'s, with the maximum reduction achieved at $n$ = $m$/2 for $m$ even, and $n$ = ($m$±1)/2 for $m$ odd. For 60° arc cells dispersion at the 2 intermediate F quads between a pair of $q$'s is reduced from the nominal matched dispersion $D_{max}$ to:

$$D_{max} \rightarrow D_{max} \cdot (1 - \tfrac{\sqrt{3}}{2} \cdot \beta_{max} \cdot q) \quad (5)$$

where the matched dispersion $D$, and beta functions are given by:

$$D = \theta L \cdot \frac{[1 \pm 1/2 \sin(\mu/2)]}{\sin^2(\mu/2)} \quad (6)$$

$$\beta = 2L \cdot \frac{[1 \pm \sin(\mu/2)]}{\sin(\mu)} \quad (7)$$

Here, $\mu$ is the phase advance per cell, $L$ is the half-cell length and $\theta$ is the total bend per half-cell. The DR arcs are comprised of 60° cells, $L = 4.4323$ m and $\theta = 0.09520$.

## RESULTS

DR lattice functions at $\gamma_t = 7.64$ and 10.03 appear in Fig 1. The unperturbed lattice is shown at the top, and the lattice with the $\gamma_t$ quads energized at the bottom. The single strength arc quads $q$ are shown as blue circles, and the double strength straight section quads are red.

Figure 2 shows how $\gamma_t$ varies as a function on the arc quadrupole integrated strengths.

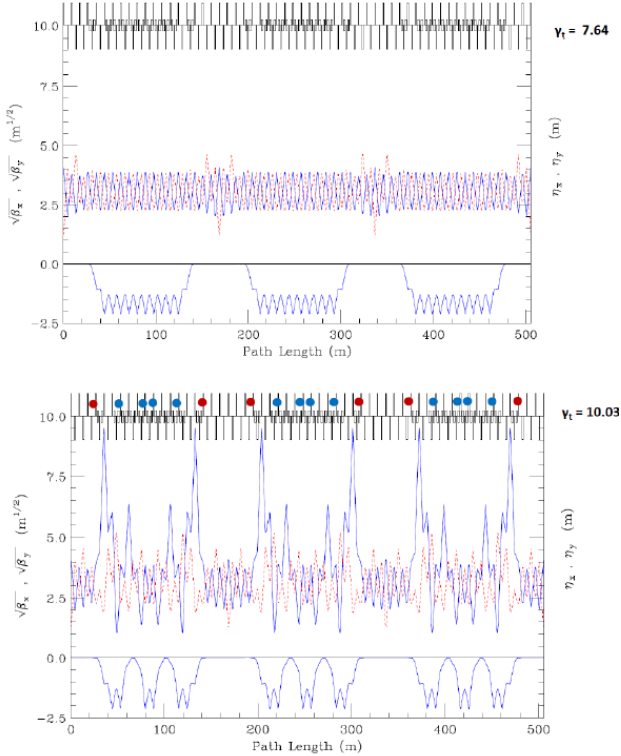

Figure 1: Lattice functions of the Delivery Ring in the unperturbed machine (top), and at injection (bottom). Transition $\gamma_t = 7.64$ and 10.04, respectively. Blue and red circles indicate sites of the arc and straight section $\gamma_t$-quads, respectively.

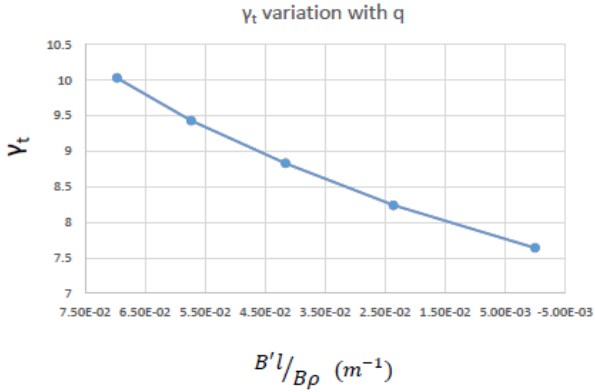

Figure 2: Variation of $\gamma_t$ with arc $\gamma_t$-quad strength.

Maximum dispersion in the arcs is unaffected by the $q$'s but the rms value is reduced from 1.42 to 1.09 m. Horizontal beta, by contrast, grows from 15.1 to 94.9 m, which should still be acceptable. For a 15 π mm-mr (95%) normalized emittance beam at 8.89 GeV/c (emittance at Booster Ring extraction), $3\sigma = 15$ mm or ~0.6 in. This is to be compared to DR quadrupole and dipole half-apertures of 1.75 in. & 1.18 in., respectively. Table 1 lists the variation of the 99% beam envelope over the entire deceleration range that the $\gamma_t$-quads are energized.

Table 1: Variation of $\gamma_t$, $\beta_{max}$, and the 15π 99% beam envelope with strength $q$ of the arc quadrupoles.

| p (GeV/c) | 8.89 | 8.33 | 7.76 | 7.20 | 6.63 | 6.07 |
|---|---|---|---|---|---|---|
| KE (GeV) | 8.00 | 7.45 | 6.88 | 6.32 | 5.76 | 5.21 |
| $\gamma_{BEAM}$ | 9.53 | 8.93 | 8.33 | 7.74 | 7.14 | 6.55 |
| $\gamma_{transition}$ | 10.03 | 9.43 | 8.83 | 7.74 | 7.64 | 7.64 |
| $\beta_{max}$ (m) | 94.9 | 72.5 | 49.5 | 30.1 | 15.1 | 15.1 |
| q (m$^{-1}$) | .0697 | .0573 | .0416 | .0236 | 0.0 | 0.0 |
| 3σ (mm) | 15.0 | 13.6 | 11.6 | 9.4 | 6.9 | 6.9 |

At injection energy and $\gamma_t = 10.03$, the arc $\gamma_t$-quads' integrated strengths are ~2.1 T-m/m (double that for the nQ07 straight section quads). These quadrupole strengths required are beyond what can be expected from trim magnets. The proposed solution is to shunt current into/around the existing magnets at these 18 locations. The nominal injection gradient of the SQC quads is 10.46 T/m. Implementing this scheme requires the nQ07, nQ13, and nQ19 gradients change to 4.26, 13.56, 13.56 T/m, respectively. In terms of current this means shunting ~140-150 A around the nQ07, while adding ~70-75 A to the nQ13, nQ19 quadrupoles.

## CONCLUSIONS

The Fermilab Delivery Ring optics can be configured to decelerate beams from the design energy of 8 GeV down to ~1-2 GeV while avoiding crossing transition. With modest modifications to the powering of select quadrupoles in the lattice, transition $\gamma_t$ can be boosted above the beam $\gamma$ for all energies. For beam $\gamma$ values that are below $\gamma_t = 7.64$ (the design transition energy value) by more than 0.5 units, the lattice optics revert to design values.

## REFERENCES

[1] REDTOP: Rare Eta Decays with a TPC for Optical Photons, http://redtop.fnal.gov/